# Unravelling the stability, electronic and physical properties in bulk and (001)-surfacesof newlysynthesized Ti$_2$ZnX (X=C, N)MAX phases

Muhammad Waqas Qureshi[a, b, *], M. A. Ali[c], Xinxin Ma[a, b, **], Guangze Tang[b], M. Usman Javed[d],

Durga Paudyal[e, f]

[a] State Key Laboratory of Advanced Welding and Joining, Harbin Institute of Technology, Harbin 150001, China

[b]School of Materials Science & Engineering, Harbin Institute of Technology, Harbin 150001, China

[c]Department of Physics, Chittagong University of Engineering and Technology (CUET), Chattogram-4349,Bangladesh

[d]Department of Physics, National Taiwan University, 10617, Taipei, Taiwan

[e] Ames Laboratory of the US DOE, Iowa State University, Ames, IA 50011, USA

[f] Electrical and Computer Engineering Department, Iowa State University, Ames, IA 50011, USA

Correspondence: [*]waqasmse@hit.edu.cn (Muhammad Waqas Qureshi), [**]maxin@hit.edu.cn (Xinxin Ma)

# Abstract

MAX phase family has been extended by the addition of late transition metals at the A-site with the expectation of diverse functional properties, such as magnetism and catalysis. Here, we present our systematic density functional investigation on the phase stability and physical properties of newly synthesized $Ti_2ZnX$ (X = C, N) phases in comparison with conventional $Ti_2AlX$ (X = C, N). Due to smaller size of N as compared to C, the unit cell dimension is reduced when C atoms are replaced by N atoms at the X-site. The thermodynamic, mechanical and dynamical stabilities are validated by estimating the formation energies, elastic constants and phonon dispersions, respectively. The elastic properties of $Ti_2ZnN$ are nearly isotropic while those of $Ti_2ZnC$ are completely anisotropic. To understand the thin-film characteristics in $Ti_2ZnX$, the surface properties with (001)-terminated slabs are investigated. Both $Ti_2ZnX$ bulk and (001)-surfaces exhibit metal-like electronic structure. There is a strong covalent bonding between Ti-X and Ti-Zn atoms. Additional states are generated at the Fermi level ($E_F$) due to the unusual *d-p* states hybridization between Ti and Zn atoms. The anisotropy in chemical bonding is confirmed by the cleavage energy difference between Ti-X and Ti-Zn atoms. Here, Ti(X)-001 and Zn-001 terminations are stable surfaces, however, in terms of chemical potentials, Zn-001 termination is the most favorable.

# 1. Introduction

A large family of ternary layered compounds called MAX phases and their two-dimensional (2D) derivatives MXenes are the materials of interest because of their remarkable properties[1–4]. The metallic-ceramic-like nature of MAX phases make them applicable in wear and corrosion protection, high-temperature structural components, protective coatings, and claddings for nuclear reactor among others[5–9]. The general formula $M_{n+1}AX_n$ (n = 1-3) represents the family to these nanolaminates where M and A denote the early transition metals and A-group elements, and X is either carbon, nitrogen or boron. The crystal structure consists the nanoscale sheets of edge shared $M_6X$ octahedra (e.g., $Ti_6C$) weakly bonded with the interleaved layers of A-group elements. The bulk MAX phases are used as precursors to produce MXenes by selective etching of A-layers because there is no direct method to their synthesis [10]. There is a constant need for the advancement of this family of nanolaminates to meet the requirement of future applications such as energy storage, supercapacitors, sensors, electromagnetic interference shielding, water purification, and photo/electrocatalysts[11–16].

An interesting aspect in MAX phases is the possibility to achieve different compositions by simply using different combinations of M, A, and X elements. To date, more than 80 different MAX phases have been synthesized experimentally while theoretical investigations are rapidly increasing[17,18]. The properties of these ternary compounds can easily be tailored by doping different M, M′; A, A′; and X, X′ elements to make quaternary MAX phase solid-solutions[19–22]. For example, the doping to Mn in $Cr_2AlC$ to form $(Cr_{1-x}Mn_x)_2AlC$ solid solution results in magnetism where ferromagnetically ordered Mn layers are exchange-coupled with non-magnetic Cr layers [23].With x > 0.6, the superconducting state of $Nb_2SC$ vanishes in solid solution $Nb_2S(B_{1-x}C_x)$ [24]. Moreover, a wide range of categories in MAX phases exist, spanning from in-plane (*i*-MAX) to out-of-plane (*o*-MAX) MAX phases[25,26] and from boride MAX to MAB phases[27,28]. The most recent advancement in the family of MAX phases is the addition of late transition noble-metal elements such as Cu, Cd, Zn, Au, Ni, Pt, and Ir in the A-site. These novel MAX phases are hard to produce from the thermodynamic equilibrium processes like hot isostatic pressing (HIP) and spark plasma sintering (SPS) because their competing M-A inter-metallic phases are more stable as compared to target MAX phase. To overcome this challenge, a top-down approach is adopted to synthesize these novel MAX phases in which A-layer in conventional MAX phases (e.g., $Ti_3SiC_2$, $Ti_3AlC_2$) is diffused out at moderate temperature and replaced by noble-metal element (e.g., Au, Zn) through a solid solution/replacement reaction[29,30].Several new MAX phases and solid-solutions such as $Ti_3AuC_2$[29], $Mo_2AuC$[31], $Ti_2CdC$ [32], $Ti_2AuN$[33], $Nb_2CuC$, $Ti_2(Al_{0.1}Cu_{0.9})N$[34], and $(Cr_{0.5}Mn_{0.5})_2AuC$[35] have been successfully synthesized. Later, the $Ti_3AuC_2$ and $Nb_2CuC$ phases have been investigated theoretically and predicted new candidates of their kind ($V_3AuC_2$ and $Cr_3AuC_2$)[36,37]. A series of Zn-based phases (Ti$_3$ZnC$_2$, Ti$_2$ZnC, Ti$_2$ZnN, and V$_2$ZnC) have been synthesized by replacement reaction in the Lewis acids ZnCl$_2$ molten salts from the conventional phases in which further exfoliation produced Cl-functional MXenes (Ti$_3$C$_2$Cl$_2$, Ti$_2$CCl$_2$)[30]. These novel phases have broadened the research interest with the expectation of diverse functional properties, such as magnetism and catalysis, due to larger *d*-electron orbitals as compared to IIIA or IVA group elements.

In the thin-film technology, MAX phases are the strong candidates for protective coatings against high-temperature oxidation, wear, corrosion, and radiation [38–41]. It is important to analyze their surface properties to understand the growth and nucleation mechanism, and early-stage oxidation. In this domain, several theoretical studies have been carried out to investigate the stable surfaces in the MAX phases where most of the work is focused on (001)-terminated surfaces because MAX phase thin-films grow along this direction [42–44]. The stability on the different slab models is evaluated by their surface energies in various chemical potentials and insightful results related to electronic properties and oxygen adsorption are discussed[45–48]. Single oxygen atom is strongly bonded onto the Al-terminated slab in Cr$_2$AlC(001) as if it is bonded to bulk fcc-Al(111)-terminated slab[48], providing the basic understanding about the early stage oxidation of Al-based MAX phases (e.g., Ti$_2$AlC, Cr$_2$AlC, Ti$_2$AlN). It is reported that the thin-film of novel Ti$_3$AuC$_2$ phase form the Ohmic contacts with SiC and are stable at $600\,°C$ for the duration of 1000 hours[29]. However, the surface investigation for these novel MAX phases is lacking and is the subject of interest here. This prospect motivates us to perform the present study with an intention to disclose the effect of Zn element in place of Al for Ti$_2$ZnC and Ti$_2$ZnN MAX compounds.

We carry out a DFT investigation for the newly-synthesized Ti$_2$ZnC and Ti$_2$ZnN phases and reveal their structural, elastic, vibrational, and electronic properties and compare them with those of Ti$_2$AlC and Ti$_2$AlN. Our results confirm that the replacement of Al by Zn does not affect in a similar way like other MAX and allowing the late transition metals may tailor and introduce new set of properties as compared to preexisting conventional MAX phases. We further study the surface stability in different environment and electronic properties of (001)-terminated slabs which disclose many interesting physical and chemical behaviour.

## 2. Computational approach

Using the framework of density functional theory (DFT), the plane wave-pseudo-potential total energy calculations were performed with generalized gradient approximation (GGA) of Perdew-Burke-Ernzerhof (PBE) [49]as the exchange-correlation (XC) functionals. According to the equation $E_{XC}^{GGA} = \int \varepsilon_{XC} F(\rho, \nabla \rho) d^3 r$, the XC is a function of charge density $\rho(r)$and spatial gradient in GGA. The position of atoms and cell parameters are optimized by using the Broyden

Fletcher-Goldfarb-Shanno (BFGS) [50] keeping the cut-off energy ($E_{cut}$) and Monk-horst-Pack meshes k-points in the Brillouin zone at 600 eV and 15×15×3, respectively. All the calculations are performed using the Cambridge Serial Total Energy Package (CASTEP) code [51]. During optimization, the self-consistent parameters i.e., maximum force, ionic displacement, stress, and total energy are fixed to 0.01 eV/Å, $5\times10^{-4}$ Å, 0.02 Pa, and $5\times10^{-6}$ eV/atom, respectively. The dynamical stability is investigated by using the finite element displacement method keeping the unit cell radius at 5 Å and 20×20×4 Monk-horst-Pack meshes k-points. For surface calculations, the relaxed unit cell is cleaved into its complementary pairs (referred to as Ti(X)-, Zn-, Ti(Zn)-, and X-terminations) by breaking the M-C and M-A bonds along the (001) direction. Surface properties are simulated by using 13 layers for Zn- and X-terminated slabs and 11 layers for Ti(X)- and Ti(Zn)-terminated slabs keeping the $E_{cut}$ at 600 eV and k-points mesh at 15×15×3. By creating a vacuum of 20 Å on both sides of the surfaces, the impact of the closest interaction between the adjacent slabs is minimized.

## 3. Results and discussion

### 3.1. Stability of bulk Ti$_2$ZnX

3.1.1. Structural and phase stability

The optimized unit cell of Ti$_2$ZnX MAX phase is depicted in Figure 1(a), which is a hexagonal symmetry having a pure Zn layer sandwiched between the edge shared [Ti$_6$X] octahedra along the *c*-axis. A total of eight atoms (Four Ti, two Zn, and two X) and two formula units are present in the unit cell and the Wyckoff positions of 4*f*(1/3, 2/3, $z_M$), 2*d* (2/3, 1/3, 1/4) and 2*a* (0, 0, 0) corresponds to Ti, Zn and X = C, N, respectively. For each composition of Ti$_2$ZnX, the plot between total energy and unit cell volume is fitted by the Birch-Murnaghan equation of state (EOS) [52] to predict the equilibrium volume (Figure 1 b, c). In Table 1, the optimized lattice parameters and unit cell volume of Zn-based phases (Ti$_2$ZnX) are presented along with those of Al-based phases (Ti$_2$AlX). It is found that the lattice parameters as well as volume of Ti$_2$ZnX is reduced when C is replaced by N because atomic radius of N (0.65 Å) is smaller than that of C (0.70 Å). Similar trend is also noticed in Ti$_2$AlX phases. Ti$_2$ZnC and Ti$_2$ZnC are most favorable with the optimized lattice parameters of *a* = 3.05Å and *c* = 13.73 Å, and *a* = 3.01Å and *c* = 13.38 Å, respectively.

The thermodynamic phase stability of bulk Ti$_2$ZnX phases is estimated by calculating their formation energy ($E_{for}$) and formation enthalpy ($\Delta H_{cp}$). Formation energy is the energy difference between the bulk Ti$_2$ZnX phase and pure comprising elements while formation enthalpy is the energy difference between the bulk Ti$_2$ZnX phase and set of competing non-MAX phase. The following equations define the $E_{for}$ and $\Delta H_{cp}$.

$$E_{for}^{Ti_2ZnX} = \frac{E_{total}^{MAX} - \left(xE_{solid}^{M} + yE_{solid}^{A} + zE_{solid}^{X}\right)}{x+y+z} \quad (1)$$

$$\Delta H_{cp} = E_{total}^{MAX} - E_{competing-phases}^{non-MAX} \quad (2)$$

where $E_{total}^{MAX}$, $E_{solid}^{M}$, $E_{solid}^{A}$, and $E_{solid}^{X}$ are the total energies of MAX phase, M, A, and X atoms in their bulk form, and $x$, $y$, $z$ are the total number of atoms in the unit cell, respectively. $E_{competing-phases}^{non-MAX}$ represents the sum of total energies of all competing non-MAX phases.

The set of competing phases are either unary, binary or ternary compound and can be determined by using the linear simplex optimization [53,54]. In Table 2, a list of all possible unary, binary or ternary compound are tabulated along with their space group, lattice parameters which are considered to calculate $\Delta H_{cp}$. Both of the MAX phases maintain the thermodynamic stability due to their negative $E_{for}$ and $\Delta H_{cp}$. Table 1 enlists the computed $E_{for}$ and $\Delta H_{cp}$ along with the set of most competing phases considered for Ti$_2$ZnX. The relation $E_{for}$ (Ti$_2$ZnN) < $E_{for}$ (Ti$_2$ZnC) suggests that the Ti$_2$ZnN phase can be synthesized with high purity as compared to Ti$_2$ZnC. Due to smaller radius of N, the required volume expansion energy of Ti to host N element in Ti$_2$ZnN is less as compared to that of Ti to host C element in Ti$_2$ZnC. Eventually, the formation energy for Ti$_2$ZnN is lower as compared to Ti$_2$ZnC.

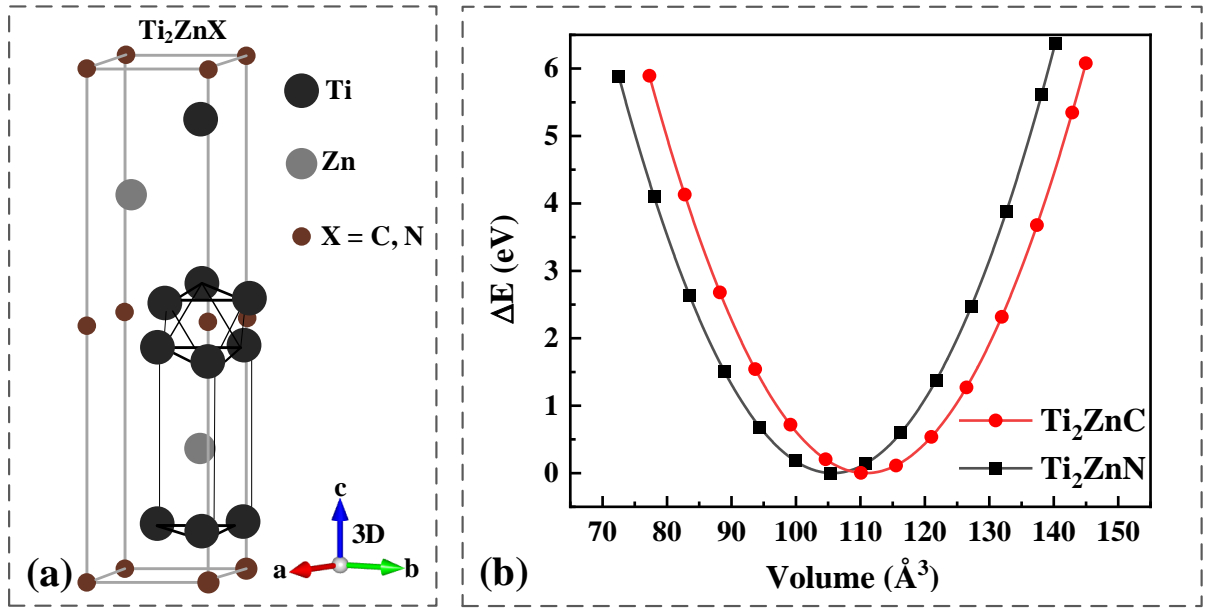

Figure 1. (a) Crystal structure of MAX phase where an Ti$_6$X octrahedra and Ti$_6$Zn trigonal prism are outlined, and (b) relative total energy ($\Delta E$) versus unit cell volume for Ti$_2$ZnX (X = C, N). $\Delta E$ is calculated by taking the ground state total energy of -10146.80 eV and -10382.79 eV as a reference for Ti$_2$ZnC and Ti$_2$ZnN, respectively.

Table 1. Ground state structure parameters, formation energy ($E_{for}$), formation enthalapy ($\Delta H_{cp}$) and set of competing phases for Ti$_2$AX phases (A = Al, Zn, X = C, N).

| Phase | $a$ (Å) | $c$ (Å) | $V$ (Å$^3$) | $z_M$ | $E_{for}$ | $\Delta H_{cp}$ | Set of competing phases |
|---|---|---|---|---|---|---|---|
| **Ti$_2$ZnC** | 3.05 | 13.73 | 110.84 | 0.0839 | -0.87 | -0.576 | Ti$_3$ZnC, Ti$_3$Zn$_3$C, Ti$_6$C$_5$ |
| **Ti$_2$ZnN** | 3.01 | 13.38 | 105.70 | 0.0838 | -1.42 | -1.096 | Ti$_3$Zn$_3$N, Ti$_2$N, TiN |
| **Ti$_2$AlC**[a] | 3.07 | 13.73 | 112.04 | 0.0834 | -0.71 | | |
| **Ti$_2$AlN**[a] | 2.99 | 13.64 | 106.06 | 0.0853 | -1.23 | | |

[a][18]

### 3.1.2. Dynamical stability

The dynamical stability of thermodynamically stable Ti$_2$ZnX phases is investigated by calculating the phonon dispersions using the finite element displacement method (Figure 2). The phonon curve is generated from second order derivative of total energy of Ti$_2$ZnX with respect to their atomic distances. There are 3 acoustic modes located at lower frequency and 21 optical modes located at higher frequency [55]. The positive phonon frequencies along with zero acoustic mode at Γ confirm the dynamical stability in both phases (Figure 2). Similar to other MAX phases, the phonon dispersions are separated by a wide gap. The lower acoustic branches are produced as a consequence of the coherent vibrations of the atoms in the lattice but not at their equilibrium positions (outside of equilibrium positions). The higher frequency modes are also separated into transverse optical (TO) and longitudinal optical (LO) modes. The optical behavior of solids is controlled by these branches. These phonon modes are produced by the out of phase movements of atoms in a lattice where one of the atoms vibrates to the left while the adjacent one moves to the right. The TO and LO frequencies are more or less constant in Ti$_2$ZnN throughout all the symmetry directions as compared to that of Ti$_2$ZnC.

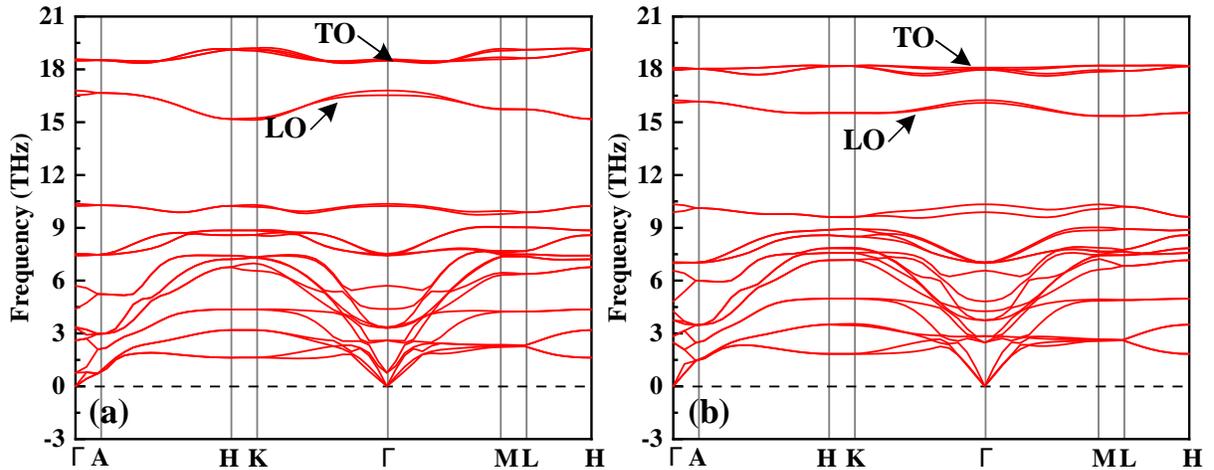

Figure 2. Phonon dispersions in (a) Ti$_2$ZnC, and (b)Ti$_2$ZnN.

### 3.1.3. Mechanical stability and elastic properties

The crystal structure and inter-atomic bonding determine the mechanical and elastic properties of a material. For hexagonal crystals of Ti$_2$ZnX, the five independent elastic constants ($C_{ij}$ = $C_{11}$, $C_{12}$, $C_{13}$, $C_{33}$, and $C_{44}$ = $C_{55}$) at 0 GPa are estimated along with elastic moduli ($B$, $G$, and $Y$) which are derived from $C_{ij}$. Table 2 lists the calculated elastic constants and properties of Ti$_2$ZnX in comparison with that of Ti$_2$AlX phases leading to following conclusive statements:

(i) Ti$_2$ZnX phases are mechanically stable as $C_{ij}$ constants satisfy the generalized Born stability criteria for hexagonal crystal, i.e., $C_{11}$ > |$C_{12}$|, $C_{44}$ > 0, $C_{66}$ > 0, ($C_{11}$+$C_{12}$) $C_{33}$ > 2$C_{13}^2$ [56,57].

(ii) $C_{ij}$ constants describe the nature of bonding in crystals. In Ti$_2$ZnX, the $C_{11}$ > $C_{33}$ represents that the bonding between the neighboring atoms along the (100) direction is comparatively stronger than that of along the (001). This leads to cleave the crystal along the (001) direction. $C_{11}$ > $C_{33}$ for Ti$_2$ZnX also indicates that the resistance to plastic deformation along the c-axis is lower as compared to the a-axis leading to elastic anisotropy. In addition, the $C_{44}$ is directly related to the hardness [58] and small $C_{44}$ for Ti$_2$ZnX corresponds to their lower hardness as compared to Ti$_2$AlX.

(iii) In Figure 3, the elastic moduli are presented in comparison with other Al-based MAX phases. Substituting the transition metal Zn at A-site lowers the elastic moduli due to difference in the bonding between M-Zn bonds and M-Al bonds. In case of Zn containing phases, bonding states between M- Zn are not occurred rather than anti-bonding states. This can be confirmed from the density of states and charge density mapping. Whereas, the M-Al atoms form bonding states among them, contribute to form covalent bonds, stronger than those of M-Zn atoms (anti-bonding states). Similar results are also reported for other MAX phases where late transition metals came as A elements such as Ti$_2$CdX (X=C and N)[59], Nb$_2$CuC [36], and Zr$_3$CdB$_4$ [60]. However, the modulus B > G is consistent for these phases as well [18]. This implies that the shear modulus limits the mechanical stability and these compounds are sensitive to deformation when subjected to shear stresses. The elastic moduli of Ti$_2$ZnX are comparable to radiation tolerant Zr$_2$AlX.

(iv) Both Ti$_2$ZnC and Ti$_2$ZnN have good resistance to elastic deformation and stiffness due to their high Young's moduli of 200 GPa and 244 GPa. Also, the nitride MAX phases have comparatively higher elastic moduli than that of carbide MAX phases which depends on [M$_6$X] blocks in their crystal structure.

(v) Ti$_2$ZnC and Ti$_2$ZnN phases with Poisson's ratios ($v$) of 0.21 and 0.19 are brittle in nature because of these smaller Poisson's ratios ($v$) as compared to the generalized value 0.26, which separates the brittleness and ductility. These phases also exhibit

(vi) covalent bonding character because of the fulfillment of the criteria of $\upsilon$ which is defined as $\upsilon \approx 0.1$, 0.25 and 0.33 for covalent, ionic and metallic bondings, respectively [61–63]. This analysis also confirms Ti$_2$ZnN is more brittle and more covalent than Ti$_2$ZnC.

(vi) The ductility index ($\mu_D = B/G$) or Pugh's ratio is another parameter to predict the ductile/brittle nature if $\mu_D$ is greater/smaller than 1.75 [64,65]. The estimated values of 1.43 and 1.32 for Ti$_2$ZnC and Ti$_2$ZnN confirming their brittleness.

(vii) The non-identical bonding nature in MAX phase crystals leads to elastic anisotropy which is estimated by the linear compressibility index, $f = k_c/k_a = (C_{11} + C_{12} - 2C_{13})/(C_{33} - C_{13})$. The computed $f > 1$ for Ti$_2$ZnC shows that this phase is elastically anisotropic and it is easily compressible along the $c$-axis than the $a$-axis because of stronger bonding along the $ab$ plane as compared to the $c$-axis [66–68]. Whereas, Ti$_2$ZnN phase is close to the isotropic limit ($f = 1$), therefore, it must exhibit identical elastic properties in all crystallographic directions.

Table 2. The calculated elastic constants ($C_{ij}$), elastic moduli ($B$, $G$, $Y$), Poisson's ratio ($\upsilon$), ductility index ($\mu_D$), linear compressibility index ($f$), and anisotropic index ($A$) for Ti$_2$ZnX phases.

| Phase | $C_{11}$ | $C_{12}$ | $C_{13}$ | $C_{33}$ | $C_{44}$ | $B$ | $G$ | $Y$ | $\upsilon$ | $\mu_D$ | $f$ | $A$ |
|---|---|---|---|---|---|---|---|---|---|---|---|---|
| **Ti$_2$ZnC** | 269.4 | 65.75 | 47.14 | 214.1 | 62.32 | 118.1 | 82.5 | 200.8 | 0.21 | 1.43 | 1.44 | 0.64 |
| **Ti$_2$ZnN** | 282.2 | 55.31 | 73.28 | 247.5 | 97.95 | 135.0 | 102.1 | 244.6 | 0.19 | 1.32 | 1.09 | 1.02 |
| **Ti$_2$AlC**[a] | 304.4 | 65.38 | 63.97 | 269.9 | 105.7 | 140.3 | 111.6 | 264.8 | 0.18 | 1.25 | 1.17 | 0.94 |
| **Ti$_2$AlN**[a] | 316.1 | 69.73 | 94.74 | 291.3 | 127.9 | 160.2 | 119.5 | 287.1 | 0.20 | 1.34 | 0.99 | 1.22 |

[a][18]

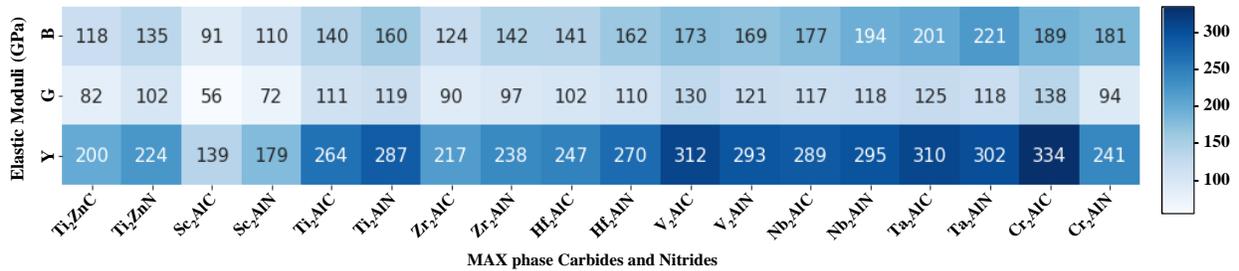

Figure 3. Comparison of elastic properties of Zn-based MAX phases with preexisting Al-based MAX phases[18].

### 3.1.4. Elastic anisotropy

Anisotropy can be used to improve the mechanical stability of materials for applications because it affects plastic deformation and the formation of micro-scale cracks. Therefore, it is necessary to study the elastic anisotropy of Ti$_2$ZnX phases due to different bonding nature caused by $C_{11} > C_{33} (C_{11} \neq C_{33})$ and linear compressibility index ($f = k_c/k_a$). The calculated anisotropic index ($A = 4C_{44}/C_{11} + C_{33} - 2C_{13}$) for Ti$_2$ZnN ≈ 1 satisfies the criteria to be isotropic whereas Ti$_2$ZnC with 0.64 exhibits anisotropic nature [69,70]. This behaviour is further investigated by visualizing the $Y$, $G$ and $v$ in directional dependent 3D plots as shown in Figure 4. The Ti$_2$ZnC phase exhibits high degree of anisotropy in $Y$, $G$ and $v$ because of its anisotropic index (< 1), which is confirmed by the non-spherical 3D visualizations (Figure a-c). The color pattern in 3D plots and stereographic projections represent the minimum and maximum values. The Young's modulus of Ti$_2$ZnC is isotropic in *a*-direction and anisotropic in *b*- and *c*-directions. Similarly, $G$ and $v$ are isotropic in *a*-direction and anisotropic in *b*- and *c*-directions with $G_{max}$ and $v_{max}$ at the 45° of *ac* and *bc* directions (Table 3). Figure 4(a′-c′) depicts the directional dependent 3D plots of Ti$_2$ZnN, which are nearly spherical due to $A$= 1.02, indicating its isotropic nature as expected. This implies that the elastic properties of Ti$_2$ZnN are expected to be identical in all crystallographic directions.

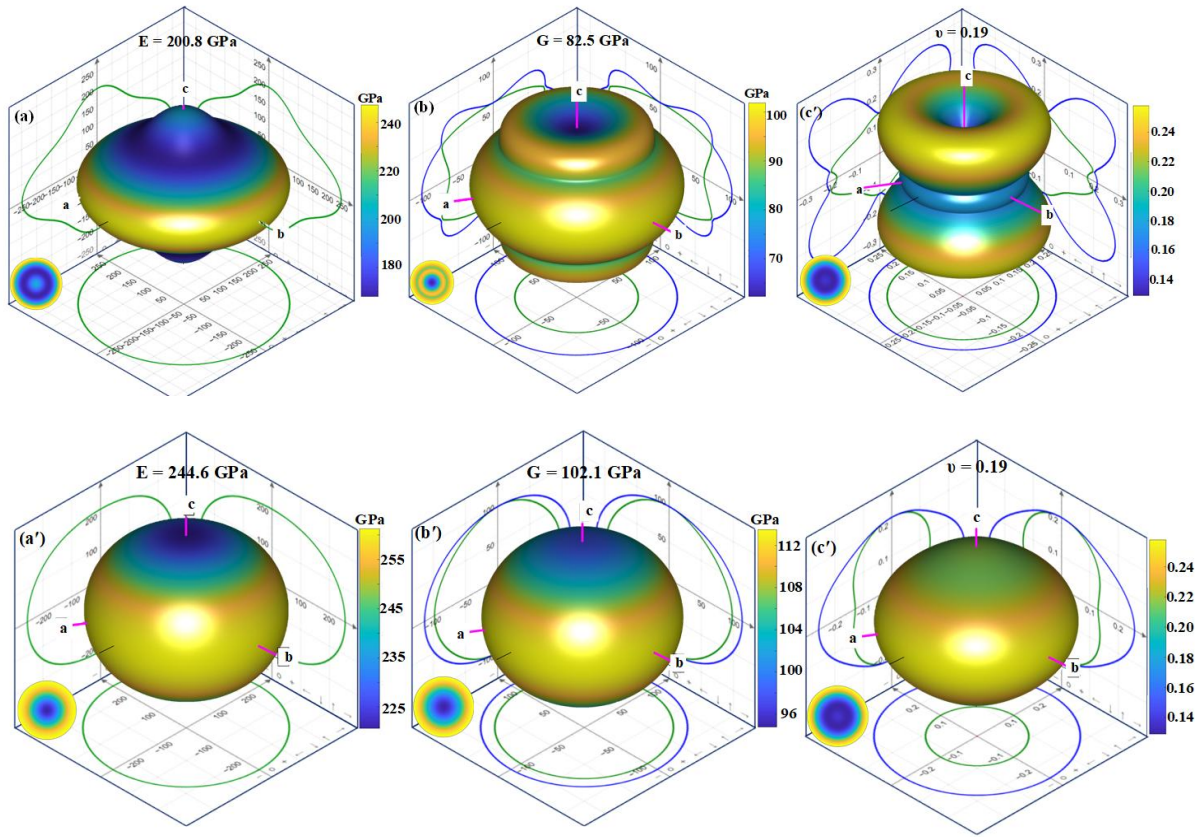

Figure 4. Demonstration of anisotropy in eleastic properties for (a-c) Ti$_2$ZnC, and (a′-c′) Ti$_2$ZnN.

Table 3. The minimum and maximum values of elestic moduli (*B*, *G*, *Y*) and their corrosponding anisotropic index in Ti$_2$ZnX.

| Phase | $Y_{min}$ | $Y_{max}$ | $A_Y$ | $G_{min}$ | $G_{max}$ | $A_G$ | $\upsilon_{min}$ | $\upsilon_{max}$ | $A\upsilon$ |
|---|---|---|---|---|---|---|---|---|---|
| Ti$_2$ZnC | 167.9 | 247.2 | 1.47 | 62.3 | 101.3 | 1.63 | 0.12 | 0.35 | 2.80 |
| Ti$_2$ZnN | 215.7 | 256.2 | 1.18 | 94.7 | 113.4 | 1.19 | 0.12 | 0.25 | 1.99 |

## 3.2. Stability of Ti$_2$ZnX (001) surfaces

3.2.1. Termination of bulk Ti$_2$ZnX into (001)-surfaces

In MAX phases, the bonding along the (001) direction is weaker than that along the (100) and the termination of bulk crystal into surfaces is favorable along the (001) direction. Experimental studies also proved that MAX phase thin-films grow along the (001) direction and single-phase epitaxial growth is also possible[42–44]. Therefore, this work is focused on (001) terminated surfaces. In this direction, termination of the bulk crystal is possible either by breaking the M-X or M-A bond. Each bond breakage creates two complementary terminations. As a result, four termination slabs are possible, i.e., X-001 and Ti(Zn)-001, and Zn-001 and Ti(X)-001by breaking the Ti-X and Ti-Zn bonds, respectively [71]. Figure 5(a-e) depicts the Ti$_2$ZnX bulk unit cell showing the Ti-X and Ti-Zn bonds and terminated slab models considered. The surface properties of Ti$_2$ZnX(001) are calculated by taking the 11 layers of Ti(X)-001 and Ti(Zn)-001, and 13 layers of Zn-001 and X-001, respectively, resulting the same atoms above and below the terminated slabs. In this way, combining each set of terminated surfaces becomes identical to stoichiometry of three bulk crystals assuring the bulk-like nature of terminated slabs. By using these surface models, the stability of (001)-surfaces is investigated from the cleavage and relaxation energies followed by surface energy in the variable chemical potentials.

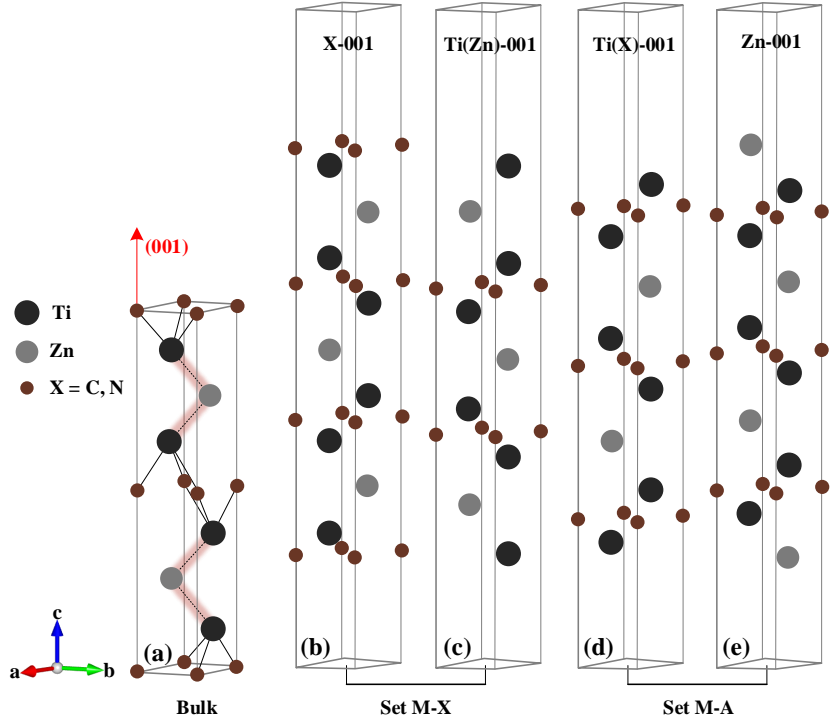

Figure 5. (a) The unit cell of bulk MAX phases indicating the M-X and M-A bonds by solid and dashed lines, respectively. Red arrow shows the (001) cleavage direction, and (b-e) four (001)-terminated slab modles considered for Ti$_2$ZnX. Each set of termination is referred to set M-X and set M-A.

The cleavage energy provides useful information related to surface properties and strength of chemical bonds. The energy released during the breaking of Ti-X and Ti-Zn bonds is referred to as cleavage energy, which is equally distributed between each set of terminated slabs:

$$E_{cleavage}^{Ti-X} = \frac{E_{unrelax}^{X} + E_{unrelax}^{Ti(Zn)} - 3E_{Ti_2ZnX}}{4A} \quad (1)$$

$$E_{cleavage}^{Ti-Zn} = \frac{E_{unrelax}^{Zn} + E_{unrelax}^{Ti(X)} - 3E_{Ti_2ZnX}}{4A} \quad (2)$$

where $E_{cleavage}^{Ti-X}$ and $E_{cleavage}^{Ti-Zn}$ are the cleavage energies of Ti-X and Ti-Zn bonds, and $E_{Ti_2ZnX}$ is the total energy of the bulk Ti$_2$ZnX phase. $E_{unrelax}^{T}$ (T =X, Ti(Zn), Ti(X), and Zn) is the total energy of terminated slab without relaxation and $A$ is the section area of the unit cell calculated by $\sqrt{3}/2a^2$.

During surface investigation, it is critical that the thickness of the surface must be sufficient to minimize the effects of adjacent terminations on each other. The convergence of cleavage energy for Ti-Zn bond was evaluated by using 9-, 13-, and 17- layers of Zn-001 termination and 7-,11-, and 15- layers for Ti(X)-001 termination, respectively. The cleavage energy difference for the termination greater than 11-layer thickness is negligible. This implies that an 11-layer thickness of a slab is sufficient enough to negate the impacts of two adjoining terminations. For reliable results, 11- and 13-layer thicknesses are considered (Table 4). For Ti$_2$ZnC, the cleavage energies for Ti-C and Ti-Zn bonds are 6.00 Jm$^{-2}$ and 1.61Jm$^{-2}$, while that of Ti-N and Ti-Zn bonds inTi$_2$ZnC are 4.91 Jm$^{-2}$ and 1.74 Jm$^{-2}$, respectively. Comparing the Ti$_2$ZnC and Ti$_2$ZnN, the breaking of Ti-C bond in Ti$_2$ZnC would require high energy as compared to that of Ti-N bond in Ti$_2$ZnN. In both compositions, the cleavage energy for M-X > M-A, indicating that M-A bond is weaker and tend to break easily when subjected to a tensile stress along the (001) direction[71,72].

Surface relaxation is also considered to calculate the surface energies of the (001)-terminations and it is described in terms of relaxation energies. The relaxation energy is defined as the difference between the total energies of the terminated slabs before and after the structural optimization:

$$E_R^T = \frac{E_{unrelax}^T - E_{relax}^T}{2A} \quad (3)$$

where $E_R^T$ represents the relaxation energy of terminated slabs $T = X$, $Ti(Zn)$, $Ti(X)$, and $Zn$. Similar to reported results, X-001 in both compositions have the maximum relaxation energies as compared to other terminations[71,72]. The N-001 termination has larger $E_R$ than that of C-001 (Table 4). The $E_R$ in Ti$_2$ZnC and Ti$_2$ZnN follows the order X-001 > Zn-001 > Ti(X)-001 > Ti(Zn)-001. Figure 6 shows the change in inter-planar distance $\Delta d_{ij}$ (where $i$ and $j$ represent the layer number in the terminated slab) during the surface relaxation. The negative values represent the reduction of inter-planar distance while the positive values represent the expansion of the inter-planar distance. The maximum change in $\Delta d_{ij}$ is found in C-001 and N-001 terminations as expected. Both C-001 and N-001 terminated surfaces display the reduction ($\Delta d_{12}$, and $\Delta d_{34}$) and expansion ($\Delta d_{23}$) in inter-planar distances. However, the change in inter-planar distance $\Delta d_{23} < \Delta d_{12}$ while change in $\Delta d_{34}$ is negligible[71,72]. The other remaining terminated surfaces follow the same trend in Ti$_2$ZnC except Ti(Zn)- and Ti(N)-terminations in Ti$_2$ZnN contradicting this trend due to negligible relaxation energy (0.001 Jm$^{-2}$).

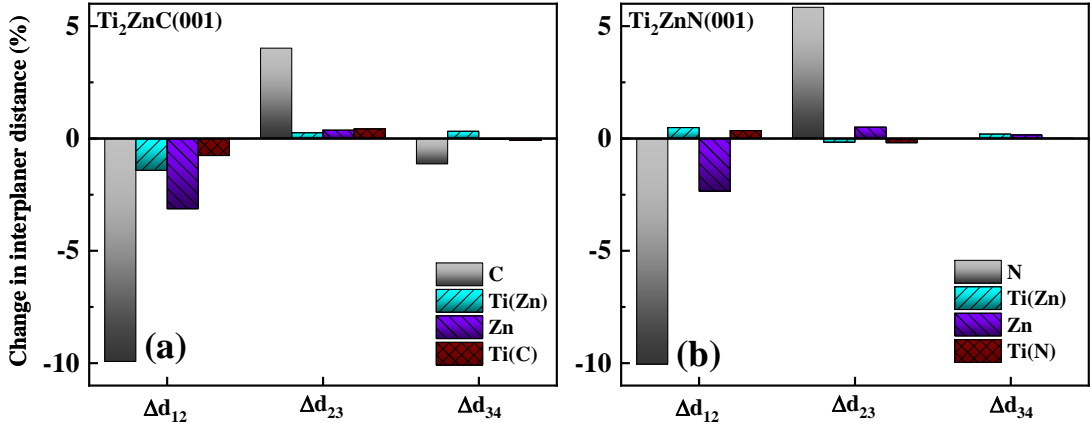

Figure 6. Relative change in interplaner distance, $\Delta d_{ij}$ (where i and j represent the layer number in the terminated slab) after relaxation of (a) Ti$_2$ZnC, and (b) Ti$_2$ZnN.

3.2.2. Surface energy

The most favorable surface configurations are determined by their lowest surface energy. The cleavage and relaxation energies are taken into account when calculating the surface energy[73]. The cleavage energy of Ti-Zn and Ti-X bonds is equally distributed into two complementary terminations, therefore, the surface energy for terminated slabs are calculated as:

$$E_S^{Zn} = E_{cleavage}^{Ti-Zn} - E_R^{Zn} \tag{4}$$

$$E_S^{Ti(X)} = E_{cleavage}^{Ti-Zn} - E_R^{Ti(X)} \tag{5}$$

$$E_S^{Ti(Zn)} = E_{cleavage}^{Ti-X} - E_R^{Ti(Zn)} \tag{6}$$

$$E_S^{X} = E_{cleavage}^{Ti-X} - E_R^{X} \tag{7}$$

where $E_S^T$ represents the surface energy of terminated slabs $T = X$, $Ti(Zn)$, $Ti(X)$, and $Zn$.

The surface energies for all terminations in Ti$_2$ZnX phases are given in Table 4. The Zn- and Ti(C)-terminated Ti$_2$ZnC(001) surfaces and Zn- and Ti(N)-terminated Ti$_2$ZnN(001) surfaces exhibit the lowest surface energies as compared to other terminations. The calculated values for these terminations are very close. In Ti$_2$ZnC, the surface energies for Zn- and Ti(C)-terminated slabs are 1.57 Jm$^{-2}$ and 1.61 Jm$^{-2}$; whereas the surface energies for C- and Ti(Zn)-terminated slabs are 4.24 Jm$^{-2}$ and 6.02 Jm$^{-2}$, respectively. The same trend is also appearing in Ti$_2$ZnN. The Zn- and Ti(N)-terminated slabs exhibit the surface energies of 1.71 Jm$^{-2}$ and 1.74 Jm$^{-2}$; while the N- and Ti(Zn)-terminated slabs exhibit the surface energies of 2.29 Jm$^{-2}$ and 4.91Jm$^{-2}$, respectively. This analysis suggests that Zn- and Ti(X)-terminations are the most favorable and stable in Ti$_2$ZnX phases, which are consistent with literature[45,46,73]. Therefore, the Zn- and Ti(X)-terminations in carbide is more stable as compared to that of nitride.

Table 4. The calculated cleavage energy, $E_{cleavage}$ (in Jm$^{-2}$), relaxation energy, $E_R$ (in Jm$^{-2}$), surface energy, $E_S$ (in Jm$^{-2}$), and change in interplanar distances, , $\Delta d_{ij}$ (in %) in Ti$_2$ZnX(001) terminated surfaces.

| *Phases* | *Bond* | *Cleavage energy* | *(001)-Termination* | *Relaxation energy* | *Surface energy* | Change in inter-planar distance | | |
|---|---|---|---|---|---|---|---|---|
| | | | | | | $\Delta d_{12}$ | $\Delta d_{23}$ | $\Delta d_{34}$ |
| Ti$_2$ZnC | Ti-C | 6.00 | C | 1.746 | 4.26 | -9.92 | 4.02 | -1.13 |
| | | | Ti(Zn) | -0.011 | 6.02 | -1.42 | 0.25 | 0.32 |
| | Ti-Zn | 1.61 | Zn | 0.039 | 1.57 | -3.13 | 0.37 | -0.04 |
| | | | Ti(C) | 0.007 | 1.61 | -0.76 | 0.42 | -0.08 |
| Ti$_2$ZnN | Ti-N | 4.91 | N | 2.627 | 2.29 | -10.04 | 5.84 | 0 |
| | | | Ti(Zn) | 0.001 | 4.91 | 0.48 | -0.16 | 0.19 |
| | Ti-Zn | 1.74 | Zn | 0.026 | 1.71 | -2.34 | 0.50 | 0.15 |
| | | | Ti(N) | 0.001 | 1.74 | 0.34 | -0.19 | 0.009 |

3.2.3. Stability analysis

The terminated surfaces are affected by the surrounding chemical environment in an open system due to the tendency of exchanging atoms. It is important to calculate the surface energy of each termination in different chemical conditions to analyze the surface stability. This approximation has been successfully used to investigate the stability of binary and ternary surfaces and interfaces[71,74–76]. Due to polar nature of Ti$_2$ZnX(001) slabs, the surface energy ($\sigma^T$) in different chemical potentials of Ti and C atoms is calculated as:

$$\sigma^T = \frac{1}{2A}\left(E_{relax}^T - N_{Ti}\mu_{Ti}^{slab} - N_{Zn}\mu_{Zn}^{slab} - N_X\mu_X^{slab} + PV - TS\right) \qquad (8)$$

where $N$ and $\mu^{slab}$ represents the total number and chemical potential of corresponding element in particular terminated slabs $T = X$, $Ti(Zn)$, $Ti(X)$, and $Zn$, respectively. The effect of *PV* and *TS* is negligible at 0 K and at normal pressure, allowing these terms to be ignored. After the optimization of terminated slabs, the chemical potentials of Ti$_2$ZnX(001) surfaces and bulk crystals are in equilibrium, i.e.,

$$\mu_{Ti_2ZnX}^{bulk} = 2\mu_{Ti}^{slab} + \mu_{Zn}^{slab} + \mu_X^{slab} \qquad (9)$$

By combining the equation (8) and (9):

$$\sigma^T = \frac{1}{2A}\left[E_{relax}^T - N_{Zn}\mu_{Ti_2ZnX}^{bulk} + (2N_{Zn} - N_{Ti})\mu_{Ti}^{slab} + (N_{Zn} - N_X)\mu_X^{slab}\right] \quad (10)$$

where $\mu^{bulk}$ indicates the chemical potential of bulk Ti$_2$ZnX phase. The accessible range of each termination can be determined by knowing the upper and lower boundary values of the three chemical potentials. The chemical potential of Ti, Zn and X atoms in Ti$_2$ZnX(001) slabs should be less than that of bulk reference phases, which gives the upper limit of chemical potentials as:

$$\Delta\mu_{Ti} = \mu_{Ti}^{slab} - \mu_{Ti}^{bulk} \leq 0 \quad (11)$$

$$\Delta\mu_{Zn} = \mu_{Zn}^{slab} - \mu_{Zn}^{bulk} \leq 0 \quad (12)$$

$$\Delta\mu_X = \mu_X^{slab} - \mu_X^{bulk} \leq 0 \quad (13)$$

where $\mu_{Ti}^{bulk}$, $\mu_{Zn}^{bulk}$, $\mu_C^{bulk}$, and $\mu_N^{bulk}$ are the chemical potentials of bulk reference phases in hcp-Ti, hcp-Zn, fcc-graphite, and N atom in $N_2$ gas molecule, respectively. Substitution of equation (11) and (13) into equation (10) gives the following expression:

$$\sigma^T = \frac{1}{2A}\left[\begin{array}{l}E_{relax}^T - N_{Zn}\mu_{MAX}^{bulk} + (2N_{Zn} - N_{Ti})\mu_{Ti}^{bulk} + (N_{Zn} - N_X)\mu_X^{bulk} \\ + (2N_{Zn} - N_{Ti})\Delta\mu_M + (N_{Zn} - N_X)\Delta\mu_X\end{array}\right] \quad (14)$$

The calculated formation energies ($\Delta H^{for}$) for bulk Ti$_2$ZnC and Ti$_2$ZnN phases are -3.504 eV and -5.701 eV, respectively, which is calculated under the constrain of following equation:

$$\mu_{Ti_2ZnX}^{bulk} = 2\mu_{Ti}^{bulk} + \mu_{Zn}^{bulk} + \mu_X^{bulk} + \Delta H_{Ti_2ZnX}^{for} \quad (15)$$

Combining the equations (9) and (15):

$$2\Delta\mu_{Ti}^{bulk} + \Delta\mu_{Zn}^{bulk} + \Delta\mu_X^{bulk} = \Delta H_{Ti_2ZnX}^{for} \quad (16)$$

Combining the equations (11-13) and (16) yields the following inequality, which provides the upper and lower limit of chemical potentials.

$$\Delta H_{Ti_2ZnX}^{for} \leq 2\Delta\mu_{Ti} + \Delta\mu_X \leq 0 \quad (17)$$

(a) Stability of Ti$_2$ZnC(001) surfaces

For Ti$_2$ZnC(001) terminated slabs, the limit of chemical potentials is:

$$-3.504 \leq 2\Delta\mu_{Ti} + \Delta\mu_C \leq 0 \quad (18)$$

By using the above range of $\Delta\mu_{Ti}$ and $\Delta\mu_C$, the stability of Ti$_2$ZnC(001) terminations as a function of $\Delta\mu_{Ti}$ and $\Delta\mu_C$ is investigated. Figure 7(a) depicts the relationship of surface energies and $\Delta\mu_{Ti}$ when $\Delta\mu_C$ is kept zero. This relationship discloses many important features. For example, the surface energy of Ti(Zn)- and C-terminated (001) slabs are independent of chemical potential of $\Delta\mu_{Ti}$ because of their constant surface energies in the whole range of $\Delta\mu_{Ti}$. Ti(Zn) has the highest surface energy confirming this termination is unfavorable and it is more reactive in the chemical potential of Ti. The Ti(C)- and Zn-terminated (001) slabs are affected by the variation of $\Delta\mu_{Ti}$. The surface energies of Ti(C)-001 decreases and Zn-001 increases with an increase in chemical potential of $\Delta\mu_{Ti}$. Over the range of $\Delta\mu_{Ti}$, Zn-001 termination has the lowest surface energy confirming it to be more stable in Ti chemical potential. The surface energies of all terminations follows the order of Ti(Zn) > Ti(C) > C > Zn at Ti-rich and Ti-poor chemical potentials. The change in surface energies with respect to $\Delta\mu_C$ is shown in Figure 7(b). It is found that the surface energies of Ti(Zn)-001 and Zn-001 increase while that of Ti(C)-001 and C-001 decrease with the increase in $\Delta\mu_C$. At low $\Delta\mu_C$, Ti(C)-001 is the least stable until $\Delta\mu_C = -2.65$ eV, otherwise, Ti(Zn)-001 becomes the least stable. The Zn-001 termination is the most favorable over the whole range of $\Delta\mu_C$. The surface energies of all terminations follow the order of Ti(C) > C > Ti(Zn) > Zn at the $\Delta\mu_C$ poor and Ti(Zn) > Ti(C) > C > Zn at the $\Delta\mu_C$ rich chemical potentials, respectively.

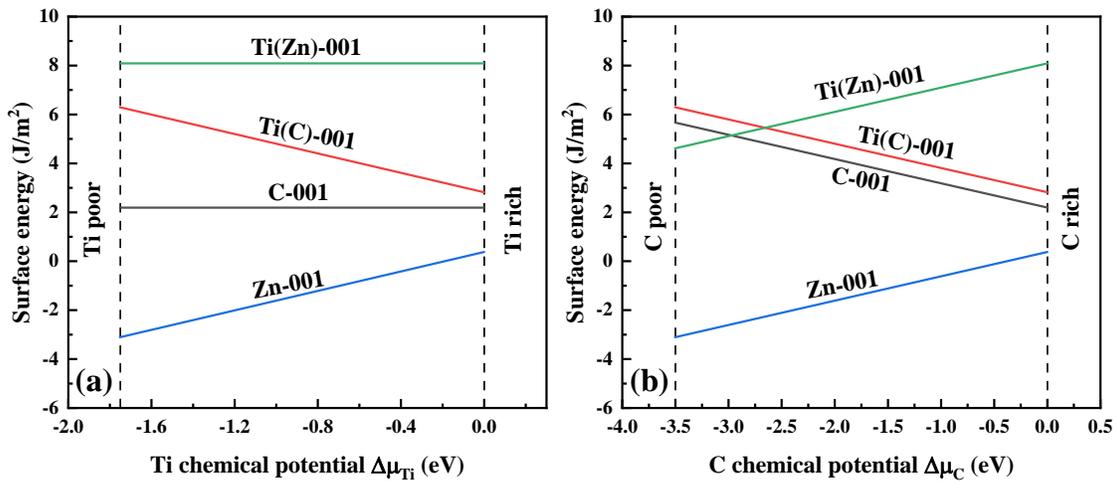

Figure 7. Stability of Ti$_2$ZnC(001) terminated surfaces in different chemical potentials of (a) $\Delta\mu_{Ti}$, and (b) $\Delta\mu_C$.

(b) Stability of Ti$_2$ZnN(001) surfaces

For Ti$_2$ZnN(001) terminated slabs, the limit of chemical potentials is:

$$-5.701 \leq 2\Delta\mu_{Ti} + \Delta\mu_N \leq 0 \tag{19}$$

By using the equation (19), the surface energies of all terminations in Ti$_2$ZnN can be calculated in relation to $\Delta\mu_{Ti}$ and $\Delta\mu_N$ (Figure 8). Similar to Ti$_2$ZnC(001) termination, the surface energies of the N-001 and Ti(Zn)-001 surfaces are independent of $\Delta\mu_{Ti}$. On the other hand, the surface energies Ti(N)-001 decrease and that of Zn-001 increase linearly with an increase in $\Delta\mu_{Ti}$. The Ti(N)-001 termination is the least at the $\Delta\mu_{Ti}$ poor potential while N-001 becomes the least stable when $\Delta\mu_{Ti}$ is higher than -0.92 eV. Therefore, the stability of Ti$_2$ZnN(001) terminations follows the order of Ti(N) > N > Ti(Zn) > Zn at low $\Delta\mu_{Ti}$ and N > Ti(N) > Ti(Zn) > Zn at high $\Delta\mu_{Ti}$ potentials, respectively. In Figure 8(b), surface energies of Ti$_2$ZnN(001) terminations are plotted as a function of $\Delta\mu_N$. The surface energies of N-001 and Ti(N)-001 decrease while that of Ti(Zn)-001 and Zn-001 increase with an increase in $\Delta\mu_N$. The stability of all terminations in the whole range of $\Delta\mu_N$ follows the order of N-001 > Ti(N)-001 > Ti(Zn)-001 > Zn-001, respectively. Over the whole range of $\Delta\mu_{Ti}$ and $\Delta\mu_N$, the Zn-001 termination is the most stable one. This agrees with the calculated surface energy of Zn-001, which is the smallest among all terminations.

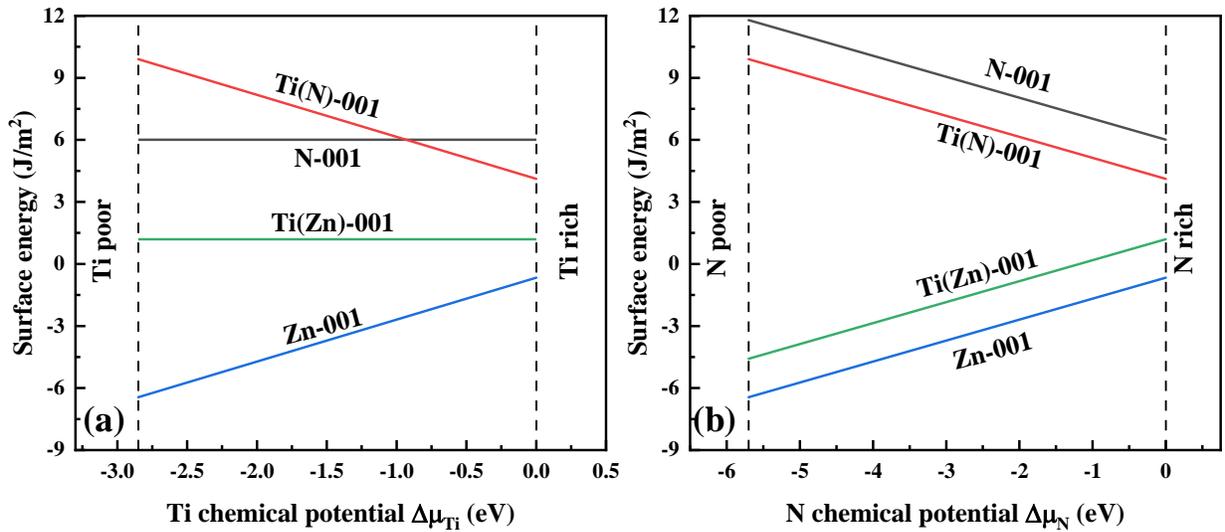

Figure 8. Stability of Ti$_2$ZnN(001) terminated surfaces in different chemical potentials of (a) $\Delta\mu_{Ti}$, and (b) $\Delta\mu_C$.

### 3.3. Electronic structure

The electronic band structure between the high-symmetry points in bulk Ti$_2$ZnX and its most favorable (001)-terminated slabs are presented in Figure 9. The ground state band structures of bulk as well as terminated surfaces exhibit the metallic characteristics due to no band gap in the vicinity of the Fermi level ($E_F$). This behaviour is due to overlapping of valance and conduction bands near $E_F$. In MAX phases, the *A-H, K-Γ, Γ-M,* and *L-H* are in-plane directions while the *Γ-A*, *H-K*, and *M-L* are in out-of-plane directions. Electronic anisotropy in band structure is due to a few or no bands crossing when $E_F$ is moving only out-of-plane directions with several bands crossing the $E_F$ in the in-plane directions. This phenomenon is responsible of non-identical electric conductivity in the *c*-axis and in the basal planes of bulk MAX phases and thin-films[77–80]. The width of conduction bands increases and unoccupied valance bands move to downwards at *Γ* (shown by black arrow) as C is replaced by N atom at X-site of Ti$_2$ZnX. The band structures of terminated surfaces also exhibit the metallic nature similar to their bulk counterparts. The numbers of bands in the surfaces are increased because each surface has one and half times as many atoms as compared to bulk counter parts. In the terminated slabs, the in-plane directions become wider while out-of-plane directions become narrower and electronic anisotropy is persistent.

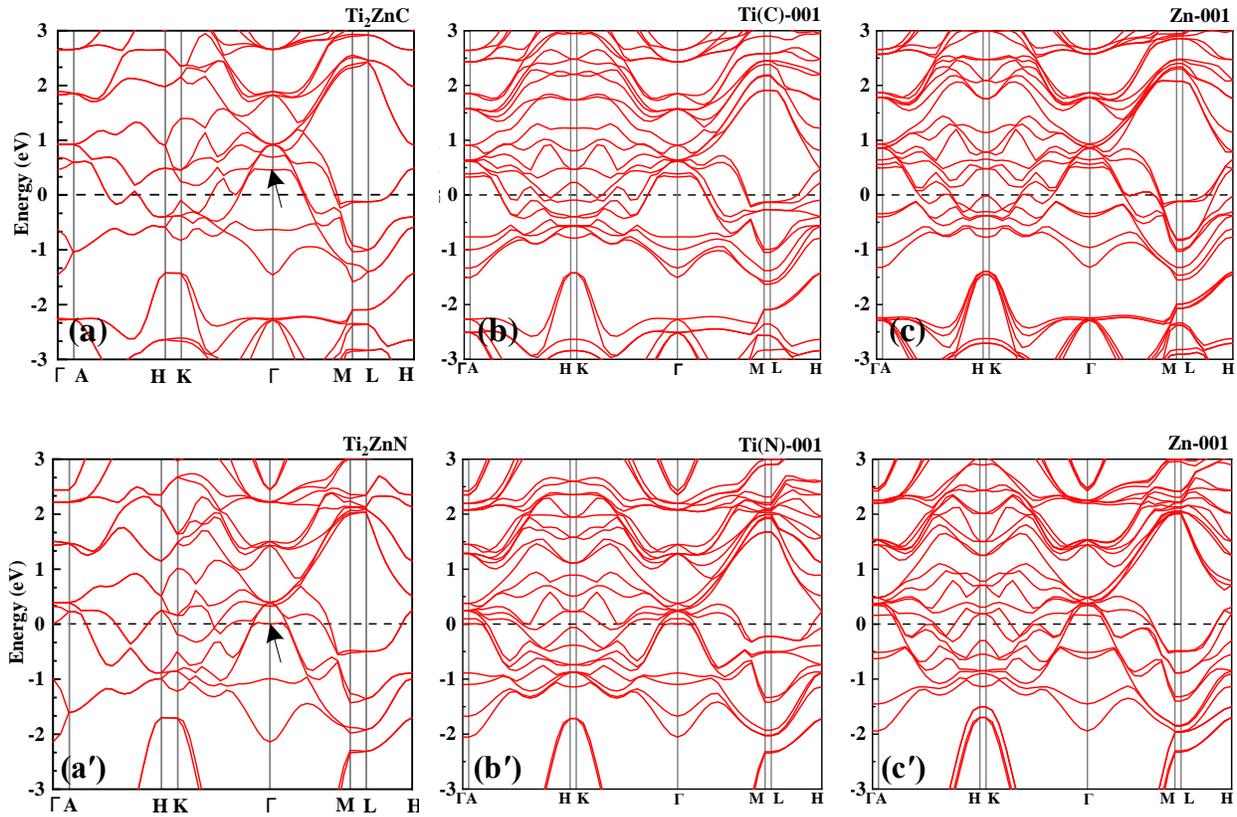

Figure 9. Electronic band structures in (a, a′) Ti$_2$ZnX bulk phases, (b, b′) Ti(X)-001 surfaces, and (c, c′) Zn-001 surfaces. $E_F$ is set to 0 eV which is represented by horizontal dashed line.

The bonding behavior of Ti$_2$ZnX phases and their (001)-terminated surfaces can be analyzed in terms of the density of states (DOS). Figure 10 displays the total and atom projected DOS in each phase where dotted and solid lines correspond to bulk and terminated surfaces, respectively. The DOS are divided into four regions and each region exhibits interaction between different sets of elements. In region 4, the total DOS is mainly contributed by the Ti-*d* and X-*s* orbitals due to strongest nearest Ti-C interactions. Due to transition element Zn at A-site in Ti$_2$ZnX phases, *d*-orbitals of Zn contribute the most to total DOS in the region 3, which is distinct from those phases having A-group elements at the A-site. In Ti$_2$ZnC, the hybridization of C-*p* and Ti-*d* is in between region 1 and 2 and hybridization between Zn-*p* and Ti-*d* is near the E$_F$ in the region 1. While in Ti$_2$ZnN, the hybridization of N-*p* with Ti-*d* and Zn-*d* is in between region 2 and 3 and the hybridization between Zn-*p* and Ti-*d* is in the region 1. This is due to the peaks of total DOS shifting towards occupied states when C is replaced by N. With this shift, the hybridization of Zn-*d* and Ti-*d* takes place in Ti$_2$ZnN, which is absent in Ti$_2$ZnC. Therefore, cleavage energy of Ti-Zn in Ti$_2$ZnN, and elastic constants and moduli are higher than those of in Ti$_2$ZnC as discussed above. At E$_F$, the finite number of total DOS indicate the metallic nature of Ti$_2$ZnX and *d* orbitals of Ti contribute the most the along with *p* orbitals of Ti and Zn, which is distinct as compared to conventional MAX phases. These orbitals are responsible for electric transport and conduction properties. The most stable (001)-terminated surfaces have similar behaviour as compared to their bulk counterparts. The most distinct feature is the higher number of states. The total DOS at $E_F$ is higher than their bulk counterparts indicating the Ti$_2$ZnX phases must exhibit metallic behaviour in bulk as well as in thin-films [71].

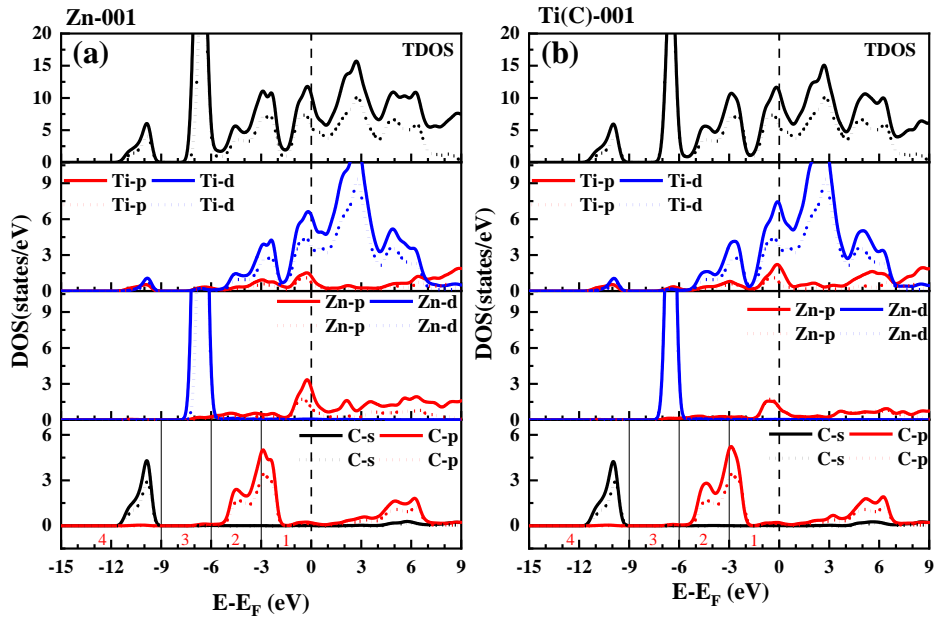

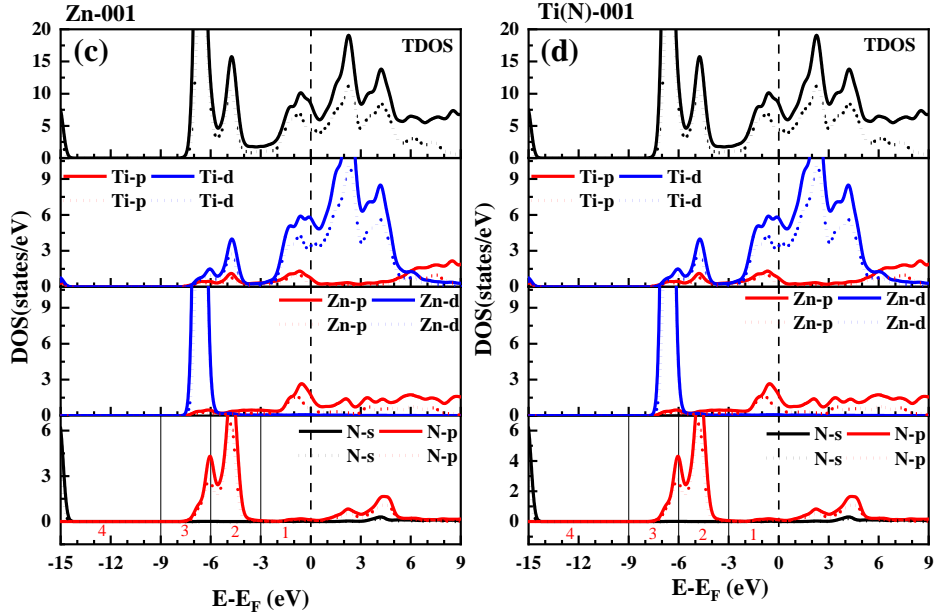

Figure 10. Electron density of states in bulk and most stable terminated surfaces (a, c) Zn-001, and (b, d) Ti(X)-001. The states in surface and bulk are indicated by the solid and dotted lines, respectively and $E_F$ is set to 0 eV.

The electron charge density provides the useful information about the chemical bonding in the unit cell and charge density map (CDM) gives its visualization. A comparison of CDM for $Ti_2ZnX$ phases with $Ti_2AlX$ and most stable surfaces in $Ti_2ZnX$(001) is given in Figure 11. The CDM for all bulk phases and surfaces are plotted along the (110). In these plots the most positive region indicates the charge accumulation responsible for covalent bonding whereas the balancing of positive and negative region indicates the charge depletion responsible for ionic bonding. The metallic bonding is represented by the most negative region. The strong charge accumulation takes place at the atomic positions of Ti, Zn and X in $Ti_2ZnX$ phases revealing the covalent nature of bonding between Ti-Zn and Ti-X (Figure 11 a, c). There is an outline of charge depletion which overlaps the nearest neighboring atoms indicating the ionic bonding which is comparatively weak. In comparison to $Ti_2AlX$ phases, the nature of bonding between Ti-X is covalent because the charge accumulation at the position Ti and X is almost similar to $Ti_2ZnX$ (Figure 11 b, d). The bonding between Ti-Al is a mixture of weak metallic and ionic, which is due to distinct charge density at the atomic position of Al[81]. Moreover, the charge between Ti and X atoms is larger than that of between Ti and Zn atoms, which supports the calculated results of cleavage energy and bond strength. The CDM of atoms in first four layers in (001)-surfaces are different from that of their bulk counterparts due to surface relaxation, which causes a great change in charge distribution in the surface area. The charge distributions of atoms below the fifth layer are similar to their bulk counterparts. The overlapping of charge density in all surfaces is increased which strengthens the chemical bonding in first four layers and causes the inward relaxation of atoms in $L_1$. This is consistent with the shortened bond length causing the reduction in inter-planar distance $\Delta d_{12}$ (Figure 6).

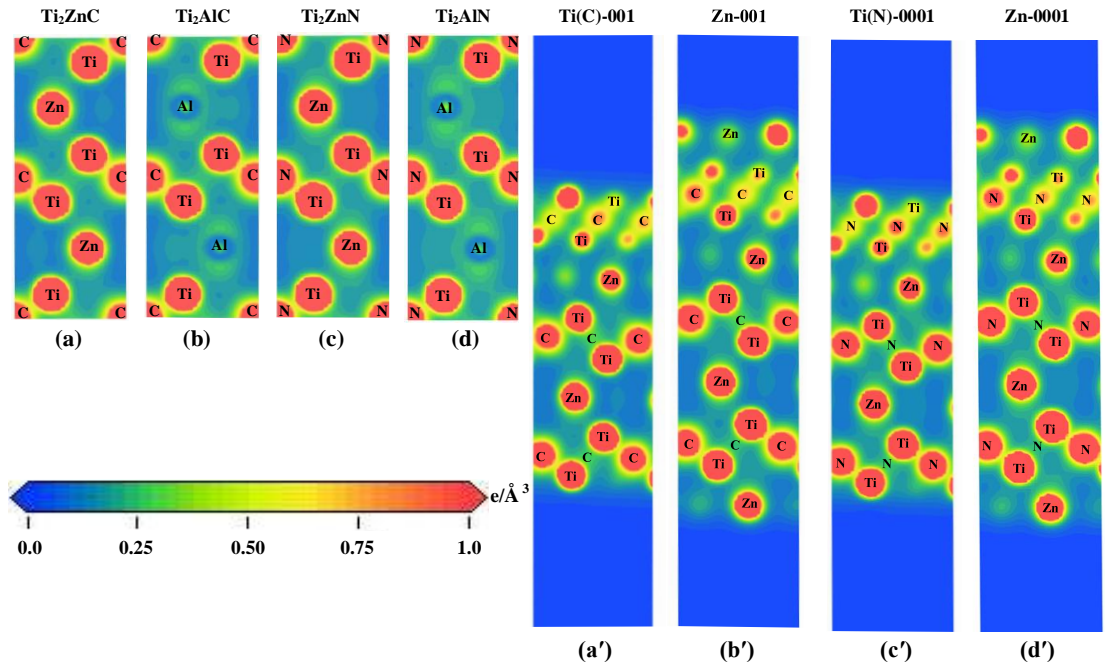

Figure 11. Electronic charge distribution alone the (110) plane in (a, c) Ti$_2$ZnX-bulk, (b, d) Ti$_2$AlX-bulk, (a′, c′) Ti(X)-001 surfaces, and (b′, d′) Zn-001 surfaces.

## 4. Conclusion

This work investigates the phase stability, electronic and physical properties of newly synthesized Ti$_2$ZnX (X = C, N) MAX phases by using the first-principles electronic structure calculations based on DFT. Screening of the phase stability from formation energy and formation enthalpy shows that the Ti$_2$ZnC and Ti$_2$ZnN phases are thermodynamically stable with the optimized lattice parameters of $a$ = 3.05Å and $c$ = 13.73 Å, and $a$ = 3.01Å and $c$ = 13.38 Å, respectively. The unit cell volume is decreased by pristine substitution of C atoms by N atoms at the X-site. The positive phonon frequencies confirm the dynamical stability of both Ti$_2$ZnC and Ti$_2$ZnN phases. The calculated elastic constants satisfy the mechanical stability criteria. Due to low shear modulus like in conventional MAX phases, Ti$_2$ZnC and Ti$_2$ZnN phases are sensitive to shear deformation. Based on elastic anisotropy factor, the elastic properties of Ti$_2$ZnC are completely anisotropic while those of Ti$_2$ZnN are nearly isotropic which is visualized by the directional dependent 3D plots.

The surface properties of Ti$_2$ZnX are studied to understand their behaviors in thin-films. By terminating the bulk unit cell, either by breaking the M-X or M-A bonds, four possible slabs (such as X-001, Ti(Zn)-001, Zn-001, and Ti(X)-001) are constructed. The cleavage energy confirms the typical anisotropic nature of chemical bonds responsible for their ceramic-metallic characteristics. The Ti-X bonds are stronger than Ti-Zn bonds, however, the weak Ti-Zn bonds limit their mechanical stability under shear stress. Zn-001 and Ti(X)-001 terminations exhibit the lowest surface energies confirming the most stable terminations. Surprisingly, Zn-001 is the only termination in both phases which is the most favorable one in the whole range of different chemical potentials i.e., $\Delta\mu_{Ti}$, $\Delta\mu_{C}$, and $\Delta\mu_{N}$. We also investigate and compare the electronic properties of bulk and the most stable (001)-terminated surfaces. There exists a covalent bond between the Ti-Zn in Ti$_2$ZnX phases as demonstrated from the charge density maps, which is distinct from Ti$_2$AlX phases where the Ti-Al bond is metallic. Additionally, the Ti-X is strongly covalent in Ti$_2$ZnX similar to Ti$_2$AlX phases supported by the cleavage energy data. Both the bulk as well as (001)-surfaces exhibit metal-like electronic structure and electronic anisotropy.

## Acknowledgement

This work is supported by the National Science and Technology Major Project (No. 2017-VII-0003-0096) and the National Natural Science Foundation of China（NSFC No.51971084 and No.52031003). A part of work done at the Ames Laboratory was conducted for the US-DOE under its contract with Iowa State University, Contract No. DE-AC02-07CH11358.